# Pinning of magnetic domain walls in multiferroics

Z.V. Gareeva[1,2], A.K. Zvezdin[1]

[1]*A.M. Prokhorov General Physics Institute, Russian Academy of Sciences, 119991 Moscow, Russia*

[2]*Institute of Molecular and Crystal Physics, Russian Academy of Sciences, 450075 Ufa, Russia*

## Abstract

The behavior of antiferromagnetic domain wall (ADW) against the background of a periodic ferroelectric domain structure has been investigated. It has been shown that the structure and the energy of ADW change due to the interaction with a ferroelectric domain structure. The ferroelectric domain boundaries play the role of pins for magnetic spins, the spin density changes in the vicinity of ferroelectric walls. The ADW energy becomes a periodical function on a coordinate which is the position of ADW relative to the ferroelectric domain structure. It has been shown that the energy of the magnetic domain wall attains minimum values when the center of the ADW coincides with the ferroelectric wall and the periodic ferroelectric structure creates periodic coercitivity for the ADW. The neighbouring equilibrium states of the ADW are separated by a finite potential barrier.

# 1 Introduction

The clamping of magnetic and electric domain walls is one of the challenges of the current research in multiferroics. Nanoscale cross – coupling between magnetic and electric order parameters has considerable potential for the technological application in data storage and spintronics, at the same time it poses a variety of fundamental physics problems. Many applications use a multi – domain configuration, therefore understanding the mechanisms allowing to control domain walls is an important issue. It is also of interest from fundamental point of view.

The pinning between ferroelectric and antiferromagnetic domains is a known phenomenon, it has been observed experimentally [1 - 10] and described theoretically [3, 4, 9 - 11] in a number of multiferroics. The origin of domain wall coupling remains under discussion. To explain the magnetoelectric pinning physical mechanisms based on the structural deformations, the piezomagnetic interaction, the Dzyaloshinskii – Moria mechanism and the non – uniform flexomagnetoelectric interaction have been proposed and discussed recently [1, 5 ,8 - 10, 15-17]. Pinned objects show peculiar properties; elucidating their behavior is of importance to understand and predict novel effects.

Herein we will focus on antiferromagnetic domain walls (ADW) in multiferroics which is one of intriguing objects of magnetism. ADW are energetically unfavorable and difficult to be visualized, but they can be stabilized by external agents: lattice strains, defects, ferroelectric domains and grain boundaries [1, 8, 11] playing the role of pinning centers.

In the present letter we consider the periodic ferroelectric domain structure as coercitivity mechanism of ADW in multiferroics. We show that the spin density distribution in ADW interacting with ferroelectric domains is changed: spins stick out the rotation plane in the vicinity of the ferroelectric boundaries. The magnetization is pinned by the ferroelectric domain walls and the pinned states are separated by a finite potential barrier. The ferroelectric domain structure gives a periodical structure of potential wells for the magnetic domain wall.

# 2 Model

As model object of the investigation we consider a $BiFeO_3$ – like multiferroic. $BiFeO_3$ has room temperature multiferrocity [18] and high value of polarization up to $10^{-4}$ C/cm$^2$ [19]. It has a rhombohedral crystalline structure and G-type antiferromagnetic ordering. The stripe – like ferroelectric domain structure is present in thin films of $BiFeO_3$ [20]. We suppose that the distribution of the polarization vector $P_z(x)$ is described by a step function, which is relevant since the width of ferroelectric domain walls (FEW) is much smaller than the width of ferroelectric domains. $OZ \parallel C_3 \parallel <111>$ (the geometry of the problem is shown in the embedment of fig.1).

We consider the properties of the ADW situated against the background of a periodic ferroelectric domain structure. The interaction between the ferroelectric domain structure and the ADW is determined by the flexomagnetoelectric interaction [16]. A key parameter for this problem appears to be the relation between the characteristic width of the ADW and the period of the ferroelectric domain structure.

In the paper [21] the specific case when the periods of the magnetic and ferroelectric domain structures coincide has been investigated. Therein we consider the more general situation unconnected with this restriction, i.e. the width of the magnetic domain wall is supposed to be comparable to or larger than the period of the ferroelectric domain structure.

The density of the free energy in terms of polar coordinates of the unit antiferromagnetic vector $\mathbf{l} = (\sin\theta\cos\varphi, \sin\theta\sin\varphi, \cos\theta)$ is written as [11]

$$f = A\left[\left(\frac{d\theta}{dx}\right)^2 + \sin^2\theta\left(\frac{d\varphi}{dx}\right)^2\right] + \left(K_1 - \frac{\chi_\perp D_1^2}{2}P_z^2(x)\right)\sin^2\theta + K_2\sin^2\theta\cos^2\varphi - D_2 P_z(x)(\cos\varphi\frac{d\theta}{dx} - \sin\theta\cos\theta\sin\varphi\frac{d\varphi}{dx}) \quad (1)$$

The free energy density (1) incorporates the energy of the non – uniform exchange interaction ($A$ is the exchange interaction constant or the exchange stiffness), the energy of the magnetic anisotropy ($K_1$, $K_2$ are the orthorhombic magnetic anisotropy constants), the energy of the uniform magnetoelectric interaction ($D_1$ is the constant of the uniform magnetoelectric interaction of Dzyaloshinskii – Moria type, $\chi_\perp$ is the magnetic susceptibility in the direction perpendicular to the antiferromagnetic vector $\mathbf{l}$) and the energy of the non – uniform magnetoelectric (flexomagnetoelectric) interaction ($D_2$ is the constant of this interaction). The later plays a central role in this theory. Minimization of the free-energy functional $F = \int f dV$ leads to the Euler - Lagrange equations

$$\frac{d^2\theta}{d\xi^2} - \left\{\left(\frac{d\varphi}{d\xi}\right)^2 - \kappa - \cos^2\varphi\right\}\sin\theta\cos\theta \qquad (2)$$
$$+ 2\varepsilon_0\varepsilon(\xi)\sin\varphi\frac{d\varphi}{d\xi}\sin^2\theta = \varepsilon(\xi)\varepsilon_0\frac{d\varepsilon}{d\xi}\cos\varphi$$

$$\frac{d}{d\xi}\left\{\sin^2\theta\frac{d\varphi}{d\xi}\right\} - \sin^2\theta\sin\varphi\cos\varphi \qquad (3)$$
$$- 2\varepsilon_0\varepsilon(\xi)\sin\varphi\frac{d\theta}{d\xi}\sin^2\theta = -\varepsilon_0\frac{d\varepsilon}{d\xi}\sin 2\theta\sin\varphi$$

where

$$\xi = \frac{x}{\Delta}, \quad \Delta = \sqrt{\frac{A}{|K_2|}}, \quad \varepsilon_0 = \frac{D_2|P_z|}{2\sqrt{A|K_2|}}, \quad \kappa = \frac{|K_{1eff}|}{|K_2|}, K_{1eff} = K_1 - \frac{\chi_\perp}{2}D_1^2P_z^2$$

$\varepsilon(\xi)$ is a periodic function with period $2d$ being equal to

$$\varepsilon(\xi) = \begin{cases} 1, & 0 < \xi < d \\ -1, & -d < \xi < 0 \end{cases}$$

on a range [ $-d$, $d$], where $d$ is the width of the ferroelectric domain.

The profile of the ADW determined by the angles $\theta$, $\varphi$ is found by the solution of eqs. (2), (3). The dependences $\theta(\xi)$, $\varphi(\xi)$ are constructed for the following values of the physical parameters $A=2\cdot10^{-7}$ erg/cm, $|K_1|=6\cdot10^6$ erg/cm$^3$, $|K_2|=1\cdot10^6$ erg/cm$^3$, $P_0=6\cdot10^{-5}$ C/cm$^2$, $\lambda=62\cdot10^{-7}$ cm, $D_2 = \frac{4\pi A}{\lambda P_0}$

($\kappa=|K_1|/|K_2|=6$, $\varepsilon_0 = \frac{D_2 P_0}{2\sqrt{A|K_2|}} = 0.45$;

$\Delta = \sqrt{\frac{A}{|K_2|}} = 4.4\cdot10^{-7}$ cm). Plots $\theta(\xi)$, $\varphi(\xi)$ are depicted in fig.1.

It is seen that the antiferromagnetic vector rotates in the *XOY* plane (fig.1 b), in the vicinity of the ferroelectric domain walls the deviations $\Delta\theta$ of the angle $\theta$ from the equilibrium state $\theta_0=\pi/2$ occur (fig. 1 (a)). Such kind of dependence $\theta(\xi)$ points out that magnetic spins interlock on the ferroelectric wall. Notice that the dependence $\theta(\xi)$ is of an asymmetric form. One can see that the position of the ferroelectric domain structure depicted in fig.1 is fixed. Hereinafter we show that the presented arrangement of ferroelectric domains is favorable from the energetic point of view. We define the ADW energy as the difference between the energy of magnetic non – uniformity and the energy of the quasi- homogeneous multiferroic magnetic state [17].

Let's introduce the parameter $\xi_0$ denoting a displacement of the ferroelectric domain structure relative to the center of the ADW. The ADW energy as a function of $\xi_0$ is written as

$$\sigma_{DW} = \int_{-\infty}^{\infty} f dx - \int_{-\infty}^{\infty} f_0 dx \qquad (4)$$

$$f_0 = A\left(\frac{d\theta}{dx}\right)^2 + \left(K_1 - \frac{\chi_\perp D_1^2}{2}P_z^2(x-x_0)\right)\sin^2\theta +$$
$$K_2\sin^2\theta - D_2P_z(x-x_0)\frac{d\theta}{dx}$$,

here $f_0$ is the energy of the quasi – homogeneous magnetic state, $f$ is determined by formula (1) with $P_z(x-x_0)$. We study the dependence $\sigma_{DW}(\xi_0)$ over a period of the ferroelectric domain structure. The shape of the ADW energy is shown in fig.2. It is seen that $\sigma_{DW}(\xi_0)$ has minimums $\sigma_{eq}$ at the points in which the center of the ADW coincides with the ferroelectric domain wall. A set of equilibrium (degenerated) states are separated by an energy barrier, the height of a barrier depends on the parameters of the material and the width of the ferroelectric domain.

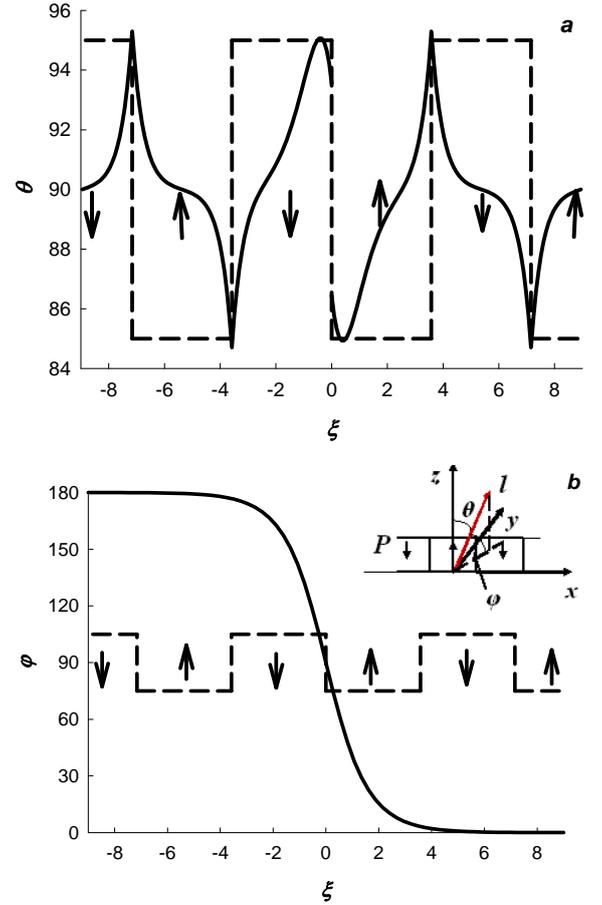

**Fig. 1** (Colour on-line) Profile of ADW along the *OX*-axis defined by the angles $\theta(\xi)$ (a), $\varphi(\xi)$ (b).

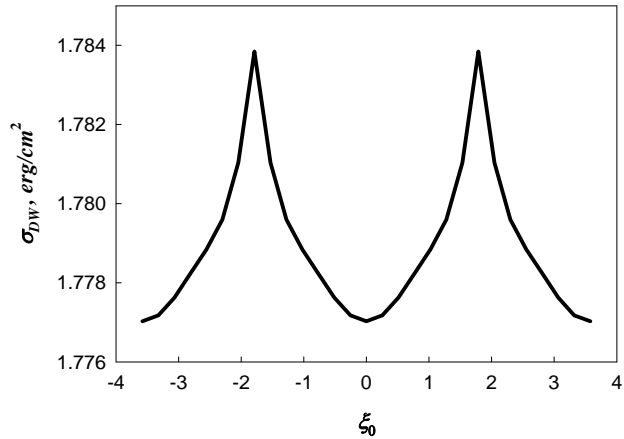

**Fig. 2** Dependence of the magnetic domain wall energy $\sigma_{DW}$ on the parameter $\xi_0$ characterizing the displacement of the magnetic domain wall relative to the ferroelectric domain structure.

## 3 Conclusion

Concluding, magnetoelectric pinning takes place in multiferroics, i.e. magnetic spins are pinned by ferroelectric domain walls. To demonstrate this property we have considered the chain of ferroelectric domains placed on the ADW. It has been shown that due to the presence of a ferroelectric structure the distribution of the spin density in the ADW changes, the spins come out from a rotation plane close to the ferroelectric domain boundaries. The domain wall energy is found to be a periodic potential on a position of the ADW relative to the ferroelectric wall. Every time the center of the ADW drops on the FEW the system comes to be in a potential well. Translation invariance changes into periodic invariance and the ADW becomes spatially confined by a potential well formed by ferroelectric domains. As a consequence, the ADW in multiferroics loses the inherited mobility and becomes coercive. The more ferroelectric domains are encountered on the ADW, the less the coercive strength. A periodic structure of ferroelectric domains creates the periodic coercitivity of ADW. This novel property of the ADW in multiferroics is of importance, it can affect the static and the dynamic behavior of domain walls.

One of the authors (AKZ) thanks A.P. Levanyuk and A.F. Popkov for discussions. The work is supported by RFBR (projects № 10.02.13302, 10.02.00846).